# Thermally Activated Plasticity in Single-Crystal Titanium: A Molecular Dynamics Study of Nanoscale Deformation

G. Markovic[1], F. J. Dominguez-Gutierrez[2]

[1]*Institute for Technology of Nuclear and Other Mineral Raw Materials, 11000 Belgrade, Serbia*
[2]*National Centre for Nuclear Research, NOMATEN CoE, ul. Andrzeja Soltana 7, 05-400 Świerk, Poland*

## Abstract

Hexagonal close-packed (hcp) titanium exhibits a complex temperature-dependent mechanical response that is central to its use in structural applications. We employ large-scale molecular dynamics simulations to investigate the nanoindentation behavior of single-crystalline α-Ti along the [0001], [10$\bar{1}$0], and [$\bar{2}$110] orientations at 10, 300, and 600 K. The simulations reveal how temperature modifies the onset of plasticity and the subsequent evolution of dislocation activity, including nucleation, glide, and the competition between basal and pyramidal ⟨c+a⟩ slip. Schmid factor mapping establishes a direct correlation between the orientation-dependent activation of slip systems and the resolved shear stress fields beneath the indenter. The results demonstrate a pronounced increase in thermally assisted dislocation motion with temperature, which manifests as diffuse slip traces and less localized pile-up patterns. Surface morphologies obtained at 300 K are consistent with atomic force microscopy observations, validating the atomistic modeling approach. At elevated temperatures, enhanced dislocation recovery and redistribution of slip pathways dominate the indentation response, highlighting the role of thermal activation in controlling plasticity in hcp titanium.

## 1. Introduction

Titanium and its alloys are materials known for their excellent balance of properties, such as high strength, relatively low density, corrosion resistance, and biocompatibility. Because of this, their applications are very broad – ranging from medical implants, through automotive and chemical industries, to the most demanding components in aerospace [1-3]. By adjusting alloy composition and applying different thermo-mechanical treatments, it is possible to significantly improve specific properties, which makes the material suitable for targeted applications [4-6]. However, in order to fully exploit titanium's potential, it is necessary to understand its fundamental deformation mechanisms. Titanium exhibits two allotropic modifications: the hexagonal close-packed (hcp) α-phase stable up to the α→β transus at about 882 °C (~1115 K), and the body-centered cubic (bcc) β-phase stable above this temperature [7]. The presence of the hcp structure in α-titanium leads to pronounced anisotropy in its mechanical behavior and a limited number of active slip systems. In terms of applications, titanium has been reported to operate in engine components at maximum service temperatures of about 400–600 °C (~673–873 K), depending on the specific alloy and design requirements [8-10]. The activation of slip systems therefore depends not only on crystallographic orientation but also strongly on temperature. These considerations also motivate the choice of representative simulation temperatures: 10 K to capture the low-temperature limit, 300 K as ambient conditions, and 600 K as a relevant high-temperature regime still within the typical use of Ti and Ti-based alloys.

Previous studies have addressed Ti from several directions. First-principles calculations have been widely used to investigate elastic properties, phase stability, and diffusion in both pure Ti and Ti-based alloys, providing important insights into their mechanical behavior at the atomic scale [11]. Molecular dynamics (MD) simulations have also been increasingly applied, modeling nanoindentation in hcp titanium and related systems to capture dislocation nucleation, propagation, and interactions under different loading conditions, as well as phase transformations such as the α→ω transition and the deformation behavior of high-temperature titanium alloys [12-15]. At the nano- and microscale, MD simulations have further revealed grain-size-dependent deformation mechanisms in polycrystalline Ti, consistent with Hall–Petch-type behavior and critical strengthening at characteristic grain sizes, providing useful insights for potential MEMS applications [16]. On the experimental side, nanoindentation and microstructural characterization techniques have revealed complex plasticity features, though often without a direct way to resolve individual slip mechanisms, which makes such studies both challenging and costly [17]. For this reason, MD simulations combined with nanoindentation modeling are a powerful approach, since they allow tracking of dislocation activity and surface evolution at the atomic level, and can be linked with Schmid factor analysis to connect crystallography with slip system activation.

This study is the first to couple atomistic nanoindentation simulations with orientation-resolved Schmid factor mapping, providing a systematic view of how slip system activation in single-crystalline α-titanium depends on temperature

## 2. Methods

Molecular dynamics simulations were performed using the Large-scale Atomic/Molecular Massively Parallel Simulator (LAMMPS) [18], which allows us to study the behavior of materials under a wide range of conditions. An embedded-atom method (EAM) potential was employed, which reproduces surface properties and mechanical behavior consistent with experimental data [19]. This is particularly important for open-boundary simulations such as nanoindentation, where material–vacuum interactions and defect nucleation induced by the indenter stress play a critical role [20]. To maintain a density of ~4.5 g/cm³, three simulation cells were constructed: ~4.25×10$^6$ Ti atoms with dimensions (45.72, 50.1, 32.76 nm) for the [0001] orientation; ~7×10$^6$ atoms with (45.86, 50.15, 53.65 nm) for [10$\bar{1}$0]; and ~7.3×10$^6$ atoms with (45.86, 61.31, 45.72 nm) for [$\bar{2}$110]. Energy minimization was performed with the FIRE 2.0 algorithm (tolerance 10$^{-6}$ eV), after which the samples were thermalized at 10, 300, and 600 K for 100 ps using a Nose–Hoover NPT thermostat (time constant 100 fs). A subsequent 10 ps relaxation eliminated artificial heat accumulation.

Each sample was divided along the z-axis. The bottom 2% of the thickness was fixed, and the next 8% served as a thermostatic layer to dissipate indentation heat. The remaining atoms formed the dynamic region. A vacuum gap of 5 nm was added above the surface [refs]. Indentation was carried out using a rigid spherical indenter with force

$$F(t) = K\,(r(t) - R)^2, \qquad (1)$$

where K = 200 eV/Å³ is the force constant and R = 12 nm is the indenter radius. The indenter center followed r(t) = (x$_0$, y$_0$, z$_0$ ± vt) with an initial offset z$_0$ = 0.5 nm. The indentation velocity was v = 20 m/s during both loading and unloading. Periodic boundary conditions were applied along x and y, and the indentation depth was limited to 4 nm to reduce boundary effects. Each simulation lasted 225 ps with a timestep of 1 fs. The load–displacement curve was obtained by plotting the indenter force against penetration depth [21-24].

Load–depth curves were computed following Pathak et al. [25, 26]:

$$p = 0.5 \frac{F}{A_p}, \qquad (2)$$

where $A_p = A_1 h + B h^2$ is the projected contact area, with h the indentation depth and $A_1$, B fitted constants. This formulation enables consistent comparison with experiments at the atomic scale. Dislocation structures were analyzed using OVITO [23] and identified by the Dislocation Extraction Algorithm (DXA) [27]. Dislocations were classified by Burgers vector as ⟨0001⟩ (c-type), 1/3⟨$\bar{1}$2$\bar{1}$0⟩ basal (a-type), 1/3⟨$\bar{1}$2$\bar{1}$3⟩ pyramidal (c+a-type), and ⟨$\bar{1}$100⟩ prismatic. To quantify orientation-dependent slip activity, atom-resolved Schmid factor (m) mapping was performed:

$$m = \cos(\phi)\cdot\cos(\lambda), \qquad (3)$$

where ϕ is the angle between the loading axis and slip plane normal n, and λ the angle between the loading axis and slip direction s. In vector form:

$$m = (\sigma\cdot n)(\sigma\cdot s), \qquad (4)$$

with σ the unit loading vector. Local crystallographic orientations were obtained from atom-based quaternions and converted to rotation matrices R, which transformed slip vectors (nc, sc) from the crystal to the sample frame. For hcp Ti, basal (0001)[$\bar{1}$2$\bar{1}$0], prismatic (10$\bar{1}$0)[$\bar{1}$2$\bar{1}$0], and pyramidal (10$\bar{1}$1)[$\bar{1}$2$\bar{1}$3] (c+a-type) systems were considered. The maximum Schmid factor at each atomic site was used for visualization, yielding maps that reveal spatially resolved slip tendencies under uniaxial indentation.

## 3. Results and Discussion

Fig. 1 presents the load–displacement (LD) curves and contact pressure for nanoindentation on the [0001] (a), [10$\bar{1}$0] (b), and [$\bar{2}$110] (c) orientations of single-crystalline HCP titanium at 10, 300, and 600 K. For clarity, results for each orientation are shown separately. Multiple independent simulations were performed at 600 K to account for statistical variations induced by thermal motion, and the results are reported as averages with error bars representing the standard deviation. For the [0001] orientation, the initial portion of the loading curve, up to approximately 1.5 nm indentation depth, exhibits a characteristic elastic response that follows the Hertzian contact model at all temperatures, followed by a distinct pop-in event marking the onset of plasticity. This pop-in corresponds to the sudden nucleation and motion of dislocations beneath the indenter, causing a sharp displacement burst at nearly constant load [28]. The unloading segment of the LD curve shows the largest residual indentation depth among the three orientations, reflecting the lowest capacity for elastic recovery, which decreases further at higher temperatures. For the [10$\bar{1}$0] and [$\bar{2}$110] orientations, pop-in events are more subdued, reflecting gradual dislocation nucleation and motion, which results in smoother displacement increases rather than sharp bursts. In addition to the pop-in, sink-in of the surrounding surface contributes to the final indentation profile. The residual indentation marks at 300 K and 600 K exhibit similar morphologies for most orientations. However, in the case of [$\bar{2}$110], the residual depth is lowest at 300 K, which can be attributed to enhanced dislocation mobility and increased elastic recovery at this intermediate temperature, leading to a stronger sink-in effect compared to 10 K and 600 K.

Contact pressure profiles show fluctuations at shallow depths (<3 nm) due to surface effects, atomic-scale roughness, and the discrete nature of atomic interactions during the initial stages of contact. Beyond ~3 nm depth, the pressure stabilizes as bulk-like deformation mechanisms dominate. An inset highlights this deeper region, revealing a progressive reduction in contact pressure with increasing temperature from 10 K to 600 K. This trend is attributed to enhanced atomic mobility and thermally activated processes, which facilitate local atomic rearrangements and reduce the resistance to indentation. Overall, at 10 K and 300 K, thermal vibrations have a negligible impact on the indentation process, whereas at 600 K, thermal fluctuations significantly affect atomic mobility and the activation of deformation mechanisms. The pressure results for all orientations follow the same trend: fluctuations near the surface gradually diminish with increasing indentation depth, while temperature-dependent reduction of contact resistance becomes apparent in the bulk-dominated region.

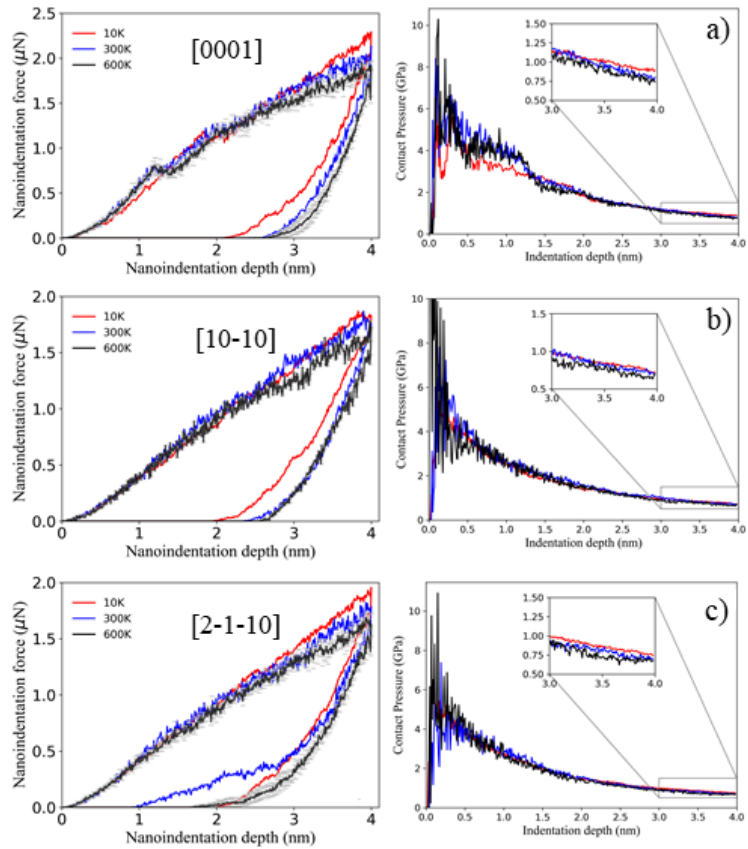

**Fig. 1.** Load–displacement (LD) curves and contact pressure during nanoindentation of single-crystalline HCP titanium at 10 K, 300 K, and 600 K for orientations [0001] (a), [10$\bar{1}$0] (b), and [$\bar{2}$110] (c). The curves show an initial Hertzian elastic response followed by pop-in events marking plasticity onset. In addition to pop-in, sink-in contributes to the residual depth, with the [$\bar{2}$110] orientation at 300 K exhibiting the lowest residual depth due to enhanced elastic recovery. Contact pressure fluctuations are strongest near the surface (<3 nm) and stabilize at larger depths, with a systematic reduction observed at higher temperatures.

3.1 Basal c-plane [0001]

During nanoindentation of single-crystalline titanium, the intense local stress beneath the indenter can induce a crystallographic phase transition from the stable hexagonal close-packed (HCP, α-phase) structure to a body-centered cubic (BCC, β-phase) configuration. This transformation occurs through a martensitic, diffusionless mechanism driven by the triaxial stress state and high strain rates, which promote collective atomic rearrangements. Crystallographically, this transition involves a collapse of the c/a ratio in the HCP lattice and shear along specific crystallographic planes, leading to the nucleation of BCC regions primarily beneath the indenter tip. The {0001} planes in HCP align with the {110} planes in BCC, while the [1$\bar{1}$20] direction in HCP corresponds approximately to the [111] direction in BCC, facilitating the transformation via minimal atomic shuffle. At elevated stress levels, transient FCC-like stacking sequences may also appear due to local distortions, although the BCC phase is more commonly stabilized under these conditions. The propensity for this transition is highest at low temperatures, where dislocation motion is suppressed, and decreases with increasing temperature as thermal activation favors conventional slip over phase transformation [22]. At 300 K, the perfect a-type dislocation with Burgers vector:

$$\frac{1}{3}<11\bar{2}0> \rightarrow \frac{1}{3}<\bar{1}0\bar{1}0> + \frac{1}{3}<01\bar{1}0>$$

was observed to dissociate into two Shockley-type partials. This dissociation produces a basal stacking-fault ribbon, stabilized by the balance between the elastic repulsion of the partials and the basal stacking-fault energy. The appearance of such partials marks the onset of thermally assisted dislocation-mediated plasticity concurrent with the stress-driven structural transition. This is reported in Fig. 2 for the maximum depth and after unloading the indenter tip at 10, 300, and 600 K.

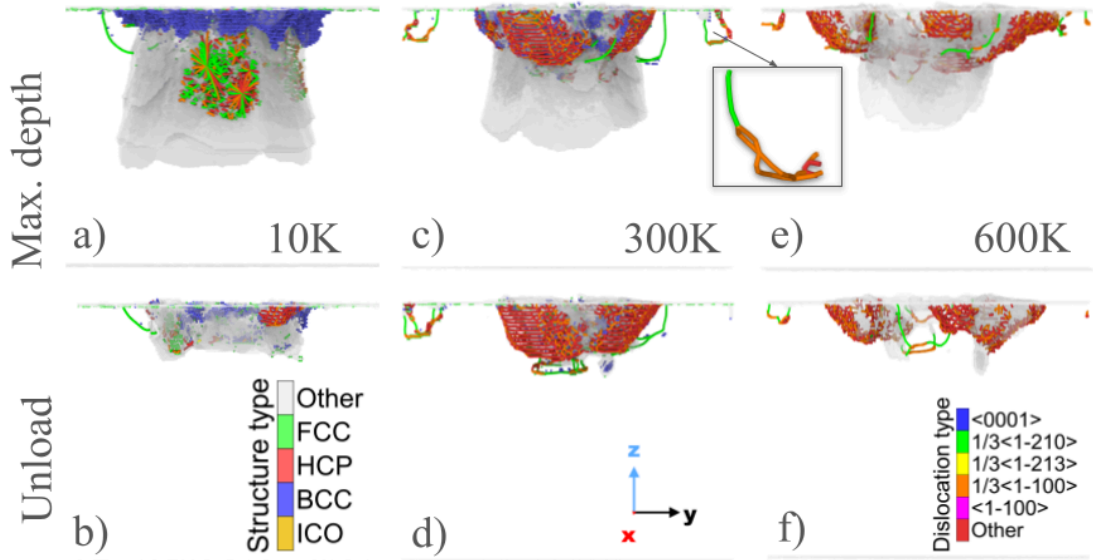

**Fig. 2**. Atomic structure and dislocation network at maximum indentation depth and after unloading, highlighting dominant basal a-type dislocations (1/3⟨$\bar{1}$210⟩) alongside c-type, prismatic, and pyramidal ⟨c+a⟩ dislocations in HCP Ti. In an inset figure, we show the dissociation of an a-type dislocation 1/3⟨11-20⟩.

Fig. 2 shows the atomic structure and dislocation network at maximum indentation depth and after unloading for different temperatures and orientations. The dominant defects are basal a-type dislocations (1/3⟨$\bar{1}$210⟩), accompanied by c-type, prismatic, and pyramidal ⟨c+a⟩ dislocations. An inset highlights the dissociation of an a-type dislocation 1/3⟨1$\bar{1}$20⟩. During nanoindentation, plastic deformation initiates once the applied load exceeds the elastic limit, which is typically marked by the pop-in event in the LD curve. Due to the highly anisotropic stress state beneath the indenter, dislocations nucleate homogeneously in subsurface regions where the resolved shear stress reaches a critical value. In HCP titanium, the main slip systems include basal {0001}⟨1$\bar{1}$20⟩, prismatic {10$\bar{1}$0}⟨1$\bar{1}$20⟩, and pyramidal {10$\bar{1}$1}⟨1$\bar{1}$23⟩ or {1$\bar{1}$22}⟨1$\bar{1}$23⟩ planes. While basal and prismatic slip generally accommodate in-plane deformation, activation of pyramidal slip is required for strain along the c-axis.

Fig. 3 presents the atomic-scale Schmid factor mapping for temperatures of 10 K, 300 K, and 600 K, illustrating the evolution of slip activity and plastic zones with increasing temperature. The Schmid factor mapping reveals a direct correlation between local crystallographic orientation and the activation of slip systems during nanoindentation. Atoms exhibiting high Schmid factor values (> 0.3) correspond to regions where the resolved shear stress on specific slip systems is maximized, indicating a higher propensity for plastic deformation. These high-value areas spatially coincide with the plastic zone beneath the indenter tip, confirming that slip is preferentially activated where the mechanical driving force is greatest [22].

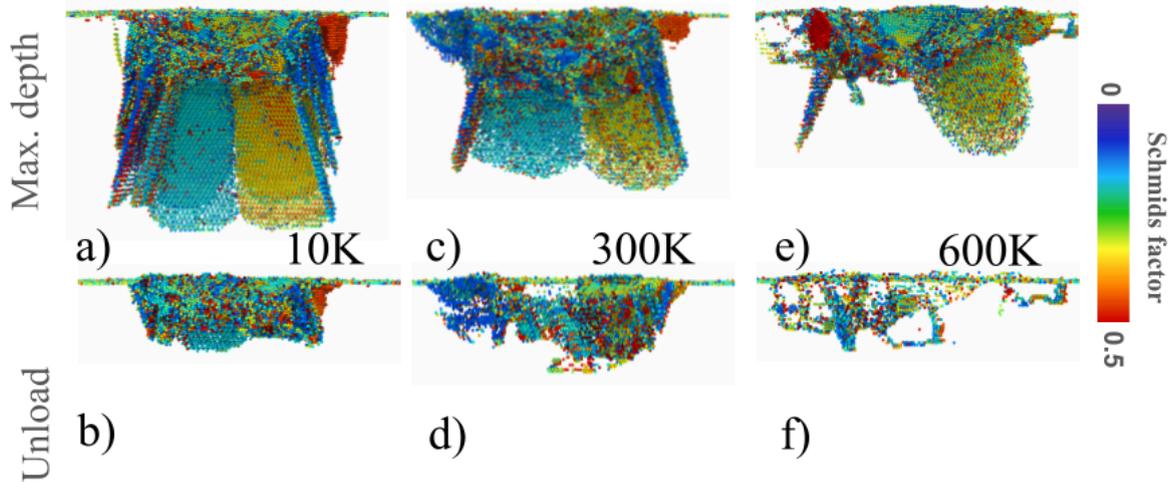

**Fig. 3**. Atomic-scale Schmid factor maps for [0001]-oriented Ti at 10 K, 300 K, and 600 K, showing the regions of high resolved shear stress corresponding to activated slip systems beneath the indenter.

Shear strain quantifies the angular distortion in a material caused by the relative sliding of atomic planes under applied stress. Unlike normal strain, which changes length along an axis, shear strain reflects the tendency of a crystal to deform by shear. In Fig. 4, atomic-scale shear strain maps are shown at maximum indentation depth and after unloading for the [0001] orientation at 10, 300, and 600 K. Shear strain is closely linked to dislocation motion, the fundamental carrier of plastic deformation. Mapping its distribution identifies subsurface regions where plasticity initiates and evolves during indentation. Analysis of the shear strain fields confirms that plastic deformation is predominantly accommodated by slip on basal (0001) planes along ⟨1$\bar{1}$20⟩ directions and on prismatic (10$\bar{1}$0) planes, with occasional activation of pyramidal (10$\bar{1}$1) slip systems under elevated stresses or at high temperature. At low temperatures, shear strain is strongly localized due to limited slip activity, while at 600 K, the deformation becomes more diffuse, consistent with thermally activated dislocation motion and the broader activation of slip systems.

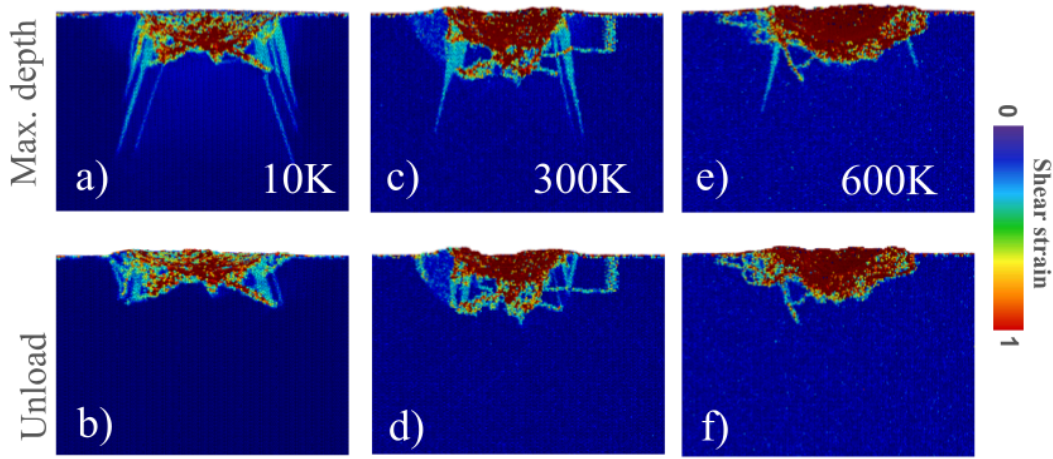

**Fig. 4.** Atomic shear strain maps of single-crystalline Ti oriented along [0001] at 10 K, 300 K, and 600 K, shown at maximum indentation depth and after unloading. The strain distribution highlights the activation of basal (0001) and prismatic (10-10) slip systems, with increased shear localization at low temperature and more diffuse deformation at elevated temperatures. Residual shear strain after unloading indicates persistent plastic deformation and dislocation structures influenced by temperature-dependent slip activity.

Fig. 5 shows the evolution of surface morphology and slip traces around the indentation site on the basal plane (0001) of single-crystalline Ti at 10, 300, and 600 K. The images correspond to maximum indentation depth and after-unloading configurations. Slip traces predominantly follow the $\{\bar{1}120\}$ planes, consistent with basal slip symmetry in HCP titanium. At 10 K, deformation is highly localized, leaving a circular residual imprint with minimal pile-up. Limited thermal activation restricts dislocation motion and plastic relaxation, so the surrounding lattice remains relatively undisturbed. At 300 K, the indentation pattern exhibits more pronounced and symmetric pile-up features, forming a quasi-hexagonal morphology around the indent. This configuration closely resembles experimental AFM observations of HCP metals at room temperature [29], reflecting active dislocation nucleation and glides along multiple in-plane directions. At 600 K, the surface morphology becomes significantly more diffuse. Enhanced atomic mobility and surface diffusion blur slip traces and partially heal surface features, reducing the residual topography after unloading.

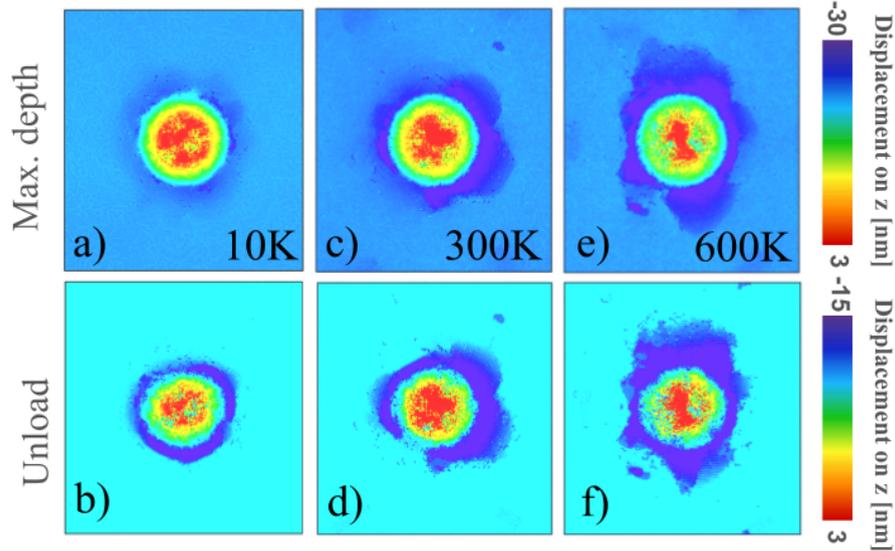

**Fig. 5.** Pile ups formation around indent, and slip traces propagation at maximum depth (upper panel) and after nanoindentation test (lower panel) at different temperatures noticing the effect of materials recovery on the material surface by different patterns.

Calculating and mapping the individual components of the strain tensor on the indented surface is essential for understanding local deformation and the resulting surface morphology, including pile-up formation and slip traces. The strain tensor captures both normal and shear deformations, which directly influence how the material's surface evolves under the indenter. Such detailed strain information helps reveal how features such as pile-up, sink-in, slip traces, and surface roughness develop in response to the complex stress state beneath the indenter. This analysis is also important for interpreting experimental surface characterization results, such as AFM images, and for validating predictive models of nanoindentation-induced plasticity [29]. Fig. 6 shows the spatial distributions of the strain tensor components $\varepsilon xx$, $\varepsilon yy$, $\varepsilon zz$, $\varepsilon xy$, $\varepsilon xz$, and $\varepsilon yz$ at 300 K. The diagonal components ($\varepsilon xx$, $\varepsilon yy$, $\varepsilon zz$) represent normal strains along the Cartesian directions and indicate local elongation or compression of the lattice. Among them, $\varepsilon zz$ contributes most strongly to the development of the pile-up pattern. The off-diagonal components ($\varepsilon xy$, $\varepsilon xz$, $\varepsilon yz$) correspond to shear strains, which describe angular distortions and relative sliding between planes. For the [0001]-oriented Ti sample, the surface morphology after nanoindentation reveals a characteristic circular pile-up pattern surrounding the indentation mark, consistent with the symmetry of basal-plane slip systems (Fig. 5). Analysis of the strain tensor components shows that $\varepsilon xz$ and $\varepsilon yz$ exhibit pronounced maximum and minimum values directly beneath the indenter tip, indicating localized shear deformation concentrated in these regions. This strain localization corresponds to the activation of slip systems that facilitate plastic flow, contributing to the observed surface pile-up morphology.

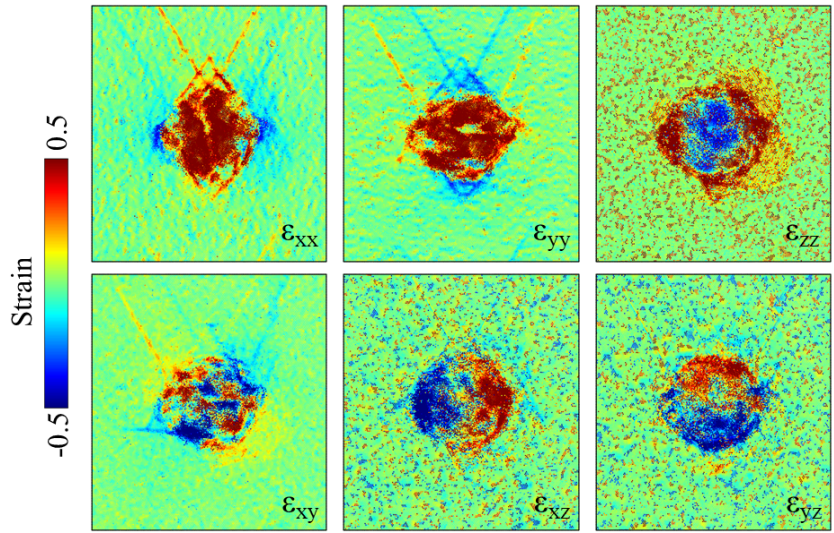

**Fig. 6**. Atomic strain tensor components for [0001]-oriented Ti after nanoindentation at 300 K.

3.2 Prismatic plane [10-10]

The $\{10\bar{1}1\}$ pyramidal planes in HCP Ti correspond crystallographically to the $\{110\}$ planes in the BCC phase, while the $\langle\bar{1}213\rangle$ (c+a) Burgers vector direction in HCP aligns approximately with the $\langle 111\rangle$ direction in BCC. Under elevated stress conditions, this crystallographic correspondence facilitates local lattice rearrangements and partial HCP→BCC transformation, while transient FCC-like stacking sequences may also appear due to distortions, although the BCC phase remains the more stable structure. The propensity for such martensitic transformation is highest at low temperatures, where dislocation mobility is limited, and decreases with temperature as thermal activation promotes conventional slip. Fig. 7 shows these transformations alongside the dislocation network at maximum indentation depth and after unloading for the prismatic orientation. The dislocation network includes c-type, basal a-type, pyramidal c+a-type, and prismatic dislocations, with loops in this orientation predominantly formed by a-type and c+a-type dislocations. At low temperature, plasticity is restricted by the high critical resolved shear stress of pyramidal systems, leading to localized dislocation activity, while increasing temperature facilitates nucleation and motion, allowing multiple slip systems to operate and generating more complex plasticity. The number of loops formed is also temperature dependent, with higher temperatures producing fewer residual loops after unloading due to enhanced annihilation and recovery. A characteristic feature of this orientation is the lasso mechanism, most evident at 300 K, which begins when stress concentrations beneath the indenter activate prismatic $\{10\bar{1}0\}\langle 1\bar{1}20\rangle$ slip systems. Dislocation segments nucleate beneath the contact area and expand radially outward, curving around the indentation axis instead of reaching the surface. Their growth and interaction cause self-intersection and reconnection, closing into prismatic loops that effectively "lasso" the stress concentration zone. At room temperature, these loops remain well defined and localized, since thermal activation improves mobility but does not yet favor extensive cross-slip or climb. They act as carriers of plasticity along prismatic planes and give rise to the characteristic two-lobe pile-up morphology observed on the $[10\bar{1}0]$ surface, while a fraction persists after unloading, leaving residual dislocation structures that contribute to slip traces and surface features.

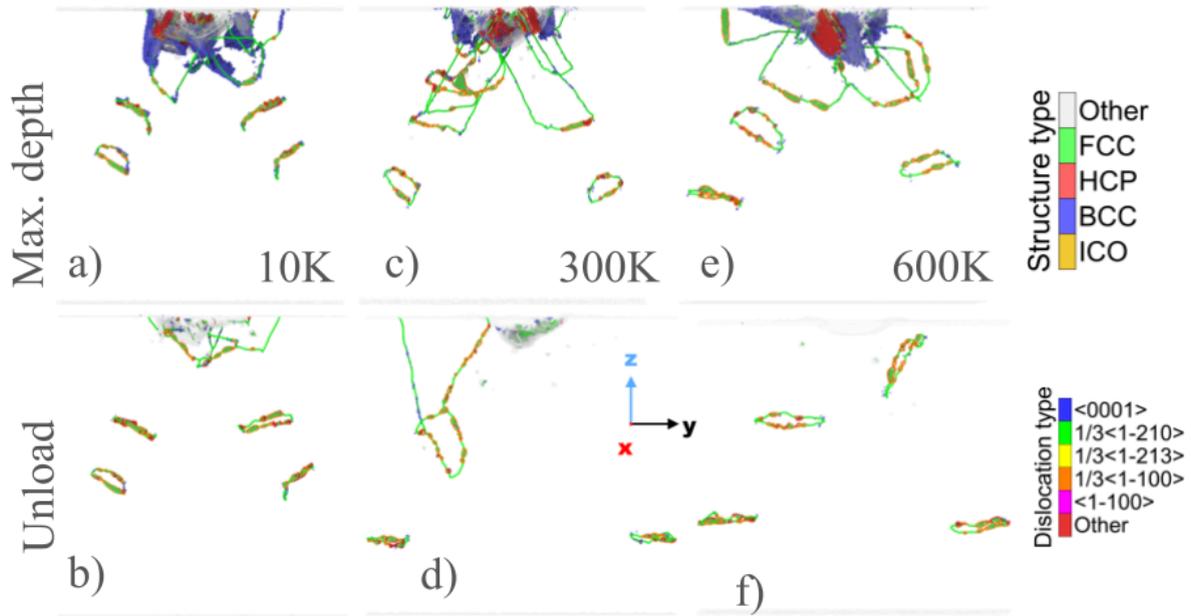

**Fig. 7.** Atomic structure and dislocation network at maximum indentation depth and after unloading of [10-10] orientation at 10, 300, and 600K.

Fig. 8 shows atomic-scale Schmid factor maps at 10, 300, and 600 K for the [10$\bar{1}$0] orientation, illustrating how slip activity and plastic zones evolve with temperature. Dislocation loops are predominantly nucleated along basal (0001) and prismatic (10$\bar{1}$0) planes, the dominant slip systems for this orientation. Atoms with high Schmid factor values correspond to regions of maximum resolved shear stress, marking the zones where plasticity initiates and where lasso processes give rise to prismatic loop formation. These regions spatially coincide with the plastic zone beneath the indenter tip, confirming that slip is activated where the mechanical driving force is highest. Increasing temperature enhances dislocation mobility and reduces energy barriers for glide, leading to more extensive plastic deformation. This analysis identifies the most favorable slip planes and directions relative to the applied load and provides atomistic insight into the anisotropic deformation behavior of [10$\bar{1}$0]-oriented Ti.

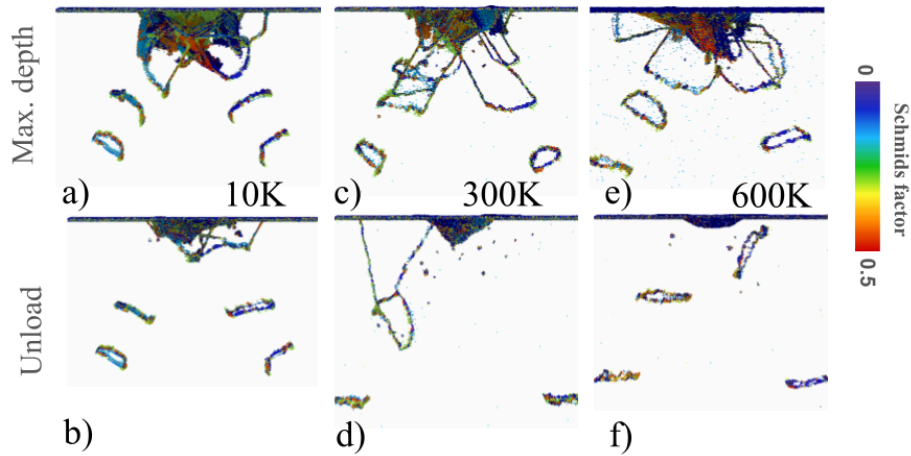

**Fig. 8.** Atomic-scale Schmid factor maps for [10-10]-oriented Ti at 10 K, 300 K, and 600 K

Fig. 9 shows atomic shear strain maps at maximum indentation depth and after unloading for [10$\bar{1}$0]-oriented Ti at 10, 300, and 600 K. With increasing temperature, shear strain magnitude rises significantly in regions near the indenter tip, reflecting enhanced atomic mobility and reduced resistance to dislocation glide. Plasticity is mainly accommodated by slip on prismatic (10$\bar{1}$0)⟨1$\bar{1}$20⟩ systems, while basal (0001) slip provides additional strain contribution. At higher stresses and temperatures, pyramidal (10$\bar{1}$1) slip systems are intermittently activated, assisting deformation and strain accommodation. This effect becomes particularly evident after unloading, when the material partially recovers and residual pyramidal slip features remain visible.

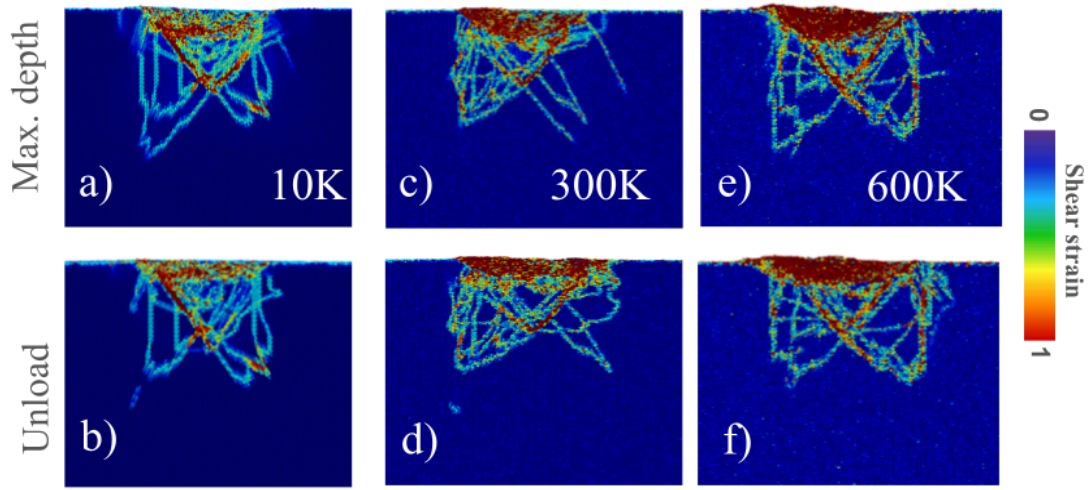

Fig. 9. Atomic shear strain maps of single-crystalline Ti oriented along [10-10] at 10 K, 300 K, and 600 K, shown at maximum indentation depth and after unloading.

Fig. 10 shows the evolution of surface morphology and slip traces around the indentation site for [$10\bar{1}0$]-oriented single-crystalline Ti at 10, 300, and 600 K. The images correspond to the configurations at maximum indentation depth and after unloading. In this orientation, slip traces are primarily associated with prismatic ($10\bar{1}0$)⟨$1\bar{1}20$⟩ and basal (0001)⟨$1\bar{1}20$⟩ slip systems. Across all temperatures, a distinctive dual-lobe morphology appears: two symmetric semi-circular features emerge on the "north" and "south" sides of the indent, aligned with crystallographic slip directions and the stress distribution of the indenter. This pattern agrees well with AFM measurements of nanoindentation in HCP Ti [29], confirming the anisotropic slip activation in this orientation. At 10 K, deformation is sharply localized, with slip traces confined near the indent edges and minimal pile-up, reflecting limited dislocation mobility. At 300 K, pile-up becomes more pronounced and symmetric, with extended slip traces indicating active dislocation glide along multiple prismatic and basal systems. At 600 K, the features become diffuse, with partially relaxed traces and reduced residual topography due to thermally activated diffusion and dislocation recovery. These trends illustrate the interplay between crystallography, temperature, and surface morphology evolution during nanoindentation in the [$10\bar{1}0$] orientation of Ti.

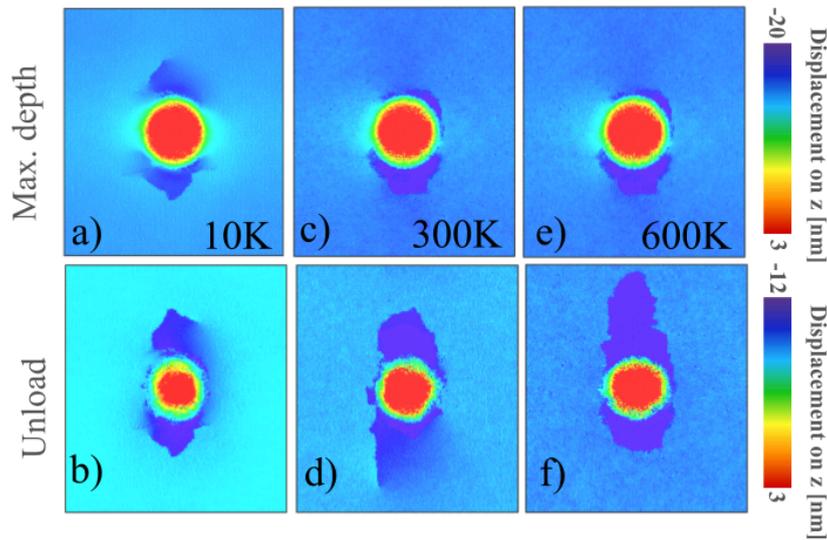

**Fig. 10.** Pile ups formation around indent for [10-10] orientation, and slip traces propagation at maximum depth (upper panel) and after nanoindentation test (lower panel) at different temperatures noticing the effect of materials recovery on the material surface by different patterns.

Fig.11 shows spatial distributions of the strain tensor components εxx, εyy, εzz, εxy, εxz, and εyz at 300 K for the [$10\bar{1}0$] orientation. Among the diagonal components, εzz contributes most to the development of pile-up features, particularly near the indentation axis. The off-diagonal components εxy, εxz, and εyz represent shear strains, describing angular distortions and relative sliding between planes. The resulting post-indentation surface morphology exhibits a distinct two-lobe pile-up pattern, aligned with the dominant slip directions and consistent with the activation of prismatic and basal slip systems in this crystallographic orientation.

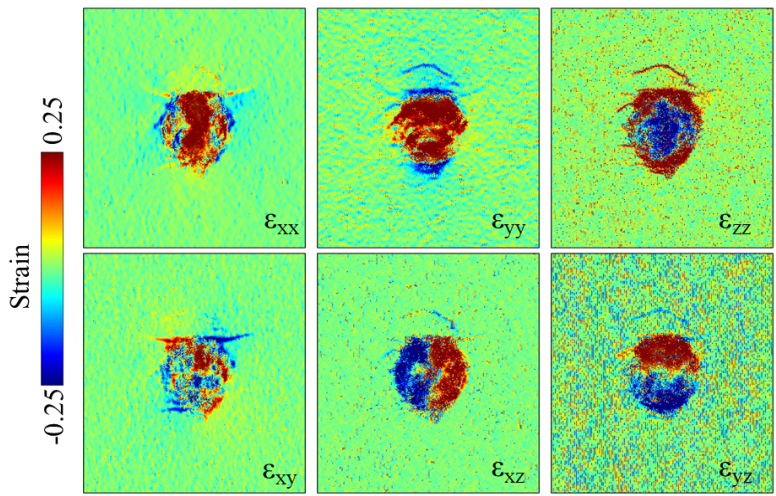

**Fig. 11.** Atomic strain tensor components for [10-10]-oriented Ti after nanoindentation at 300 K.

3.3 a-plane [21-10]

Fig. 12 shows the dislocation network at maximum indentation depth and after unloading for [$\bar{2}110$]-oriented Ti at 10, 300, and 600 K. Dislocation activity is primarily governed by the nucleation and expansion of $1/3\langle\bar{1}210\rangle$ (a-type) and $1/3\langle\bar{1}213\rangle$ (c+a-type) dislocations, as in the other orientations. While in the [$10\bar{1}0$] orientation many prismatic loops form via the lasso mechanism, in [$\bar{2}110$] loops nucleate directly beneath the indenter and subsequently expand through glide and self-interaction on multiple slip planes. Intense stress concentrations beneath the indenter activate several non-basal slip systems, and the interaction of dislocation segments with each other and with the free surface leads to the formation of closed-loop structures. These loops are frequently stabilized by lattice friction and local atomic rearrangements, including FCC-like stacking faults enclosed by the dislocation loops. At low temperature, activation of non-basal systems is limited, producing localized and directional dislocation motion. With increasing temperature, thermal energy promotes nucleation of additional dislocations, facilitates cross-slip and loop expansion, and enhances recovery and annihilation, resulting in fewer residual loops after unloading. After the indenter is removed, atoms in the plastically deformed zone largely recover the HCP structure, leaving dislocation loops as the dominant residual defects.

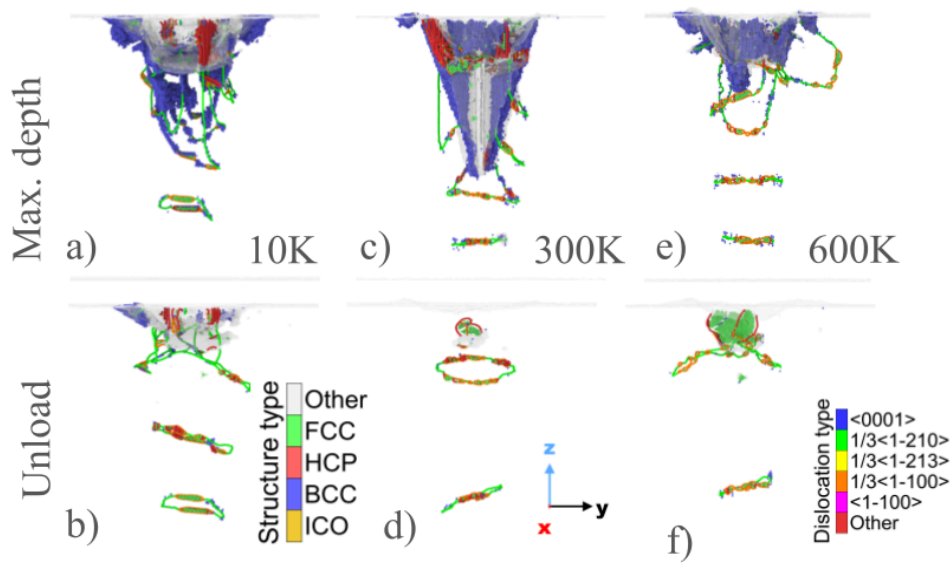

**Fig. 12**. Atomic structure and dislocation network at maximum indentation depth and after unloading of [21-10] orienation at 10, 300, and 600K

Fig. 13 presents the atomic-scale Schmid factor mapping for temperatures of 10 K, 300 K, and 600 K. For the [$\bar{2}110$] orientation, dislocation loops are predominantly nucleated along the basal {0001}, prismatic {$\bar{1}100$}, and pyramidal {$\bar{1}101$} slip planes, which are favorably oriented for shear under this loading direction. Atoms exhibiting high Schmid factor values are clustered beneath the indenter tip, coinciding with regions of maximum resolved shear stress, and marking the zones of preferred plastic deformation. Similar to the [$\bar{1}010$] orientation, loop formation is evident. This mechanism leads to the formation of complex loop structures, frequently associated with local stacking faults or FCC-like planar rearrangements. As temperature increases, the spatial extent of regions with high Schmid factor also expands, and the mobility of dislocations is enhanced through thermal activation. This results in broader and more diffuse plastic zones, increased dislocation cross-slip, and partial recovery during unloading, which collectively reduce the number of residual dislocation structures at higher temperatures. The Schmid factor mapping thus highlights how temperature and crystallographic orientation jointly govern the activation of slip systems and the anisotropic plastic response of Ti during nanoindentation.

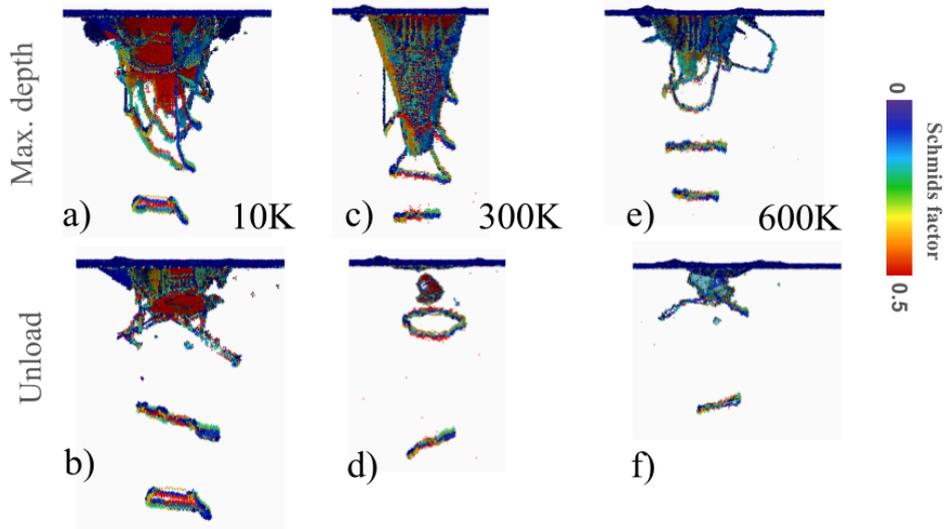

**Fig. 13**. Atomic-scale Schmid factor maps for [21-10]-oriented Ti at 10 K, 300 K, and 600 K.

Fig. 14 shows atomic shear strain maps at maximum indentation depth and after unloading for [$\bar{2}110$]-oriented Ti at 10, 300, and 600 K. With increasing temperature, a pronounced intensification of localized shear strain develops beneath the indenter, reflecting enhanced atomic mobility and reduced barriers for dislocation glide. This accumulation of shear strain indicates the activation of thermally assisted plastic mechanisms, which broaden and diffuse the deformation zones at elevated temperatures. For this orientation, plasticity is primarily governed by prismatic {$\bar{1}100$}⟨$1\bar{1}20$⟩ slip, with basal {0001} systems contributing to strain accommodation. At higher stresses and temperatures, pyramidal {$\bar{1}101$} slip systems are also activated, particularly during unloading, where recovery is partially mediated by out-of-plane dislocation motion and rearrangement.

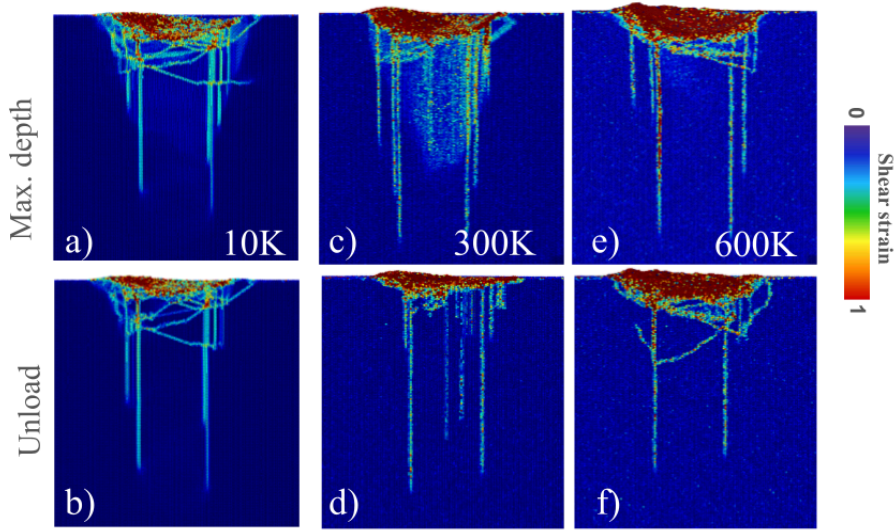

**Fig. 14.** Atomic shear strain maps of single-crystalline Ti oriented along [21-10] at 10 K, 300 K, and 600 K, shown at maximum indentation depth and after unloading.

Fig. 15 presents the evolution of surface morphology and slip trace development around the indentation site on the prismatic plane [2̄110] of single-crystalline titanium at 10 K, 300 K, and 600 K. A distinct surface morphology emerges across all temperatures, characterized by asymmetric, curved pile-up features surrounding the indent, with more diffuse lobes extending laterally rather than symmetrically as in the [101̄0] orientation. This morphology reflects the complex interaction between resolved shear stresses and crystallographic constraints inherent to the [2̄110] orientation. At 10 K, deformation remains highly localized, with sharp slip traces and minimal surface uplift, indicating limited dislocation mobility. At 300 K, the surface features become more pronounced, with visible slip lines radiating from the indent and increased pile-up suggesting enhanced dislocation activity. By 600 K, the slip traces become broader and less defined due to thermally activated recovery and possible surface diffusion effects. The progressive change in surface morphology and trace geometry underscores the temperature sensitivity of slip system activation and the anisotropic mechanical response of Ti during nanoindentation along the [2̄110] direction.

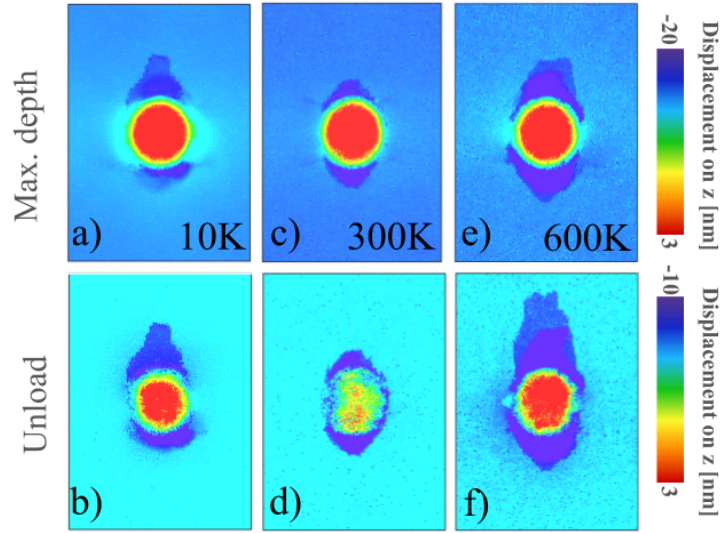

**Fig. 15**. Pile ups formation around indent, and slip traces propagation at maximum depth (upper panel) and after nanoindentation test (lower panel) at different temperatures noticing the effect of materials recovery on the material surface by different patterns.

Fig. 16 shows spatial distributions of the strain tensor components εxx, εyy, εzz, εxy, εxz, and εyz at 300 K for the [2̄110] orientation. The off-diagonal components (εxy, εxz, εyz) capture shear strains, describing angular distortions and relative sliding between crystallographic planes. Following indentation, the surface morphology exhibits a distinctly asymmetric pile-up distribution compared to other orientations, reflecting the complex interplay of active slip systems. This asymmetry is consistent with the combined activity of basal (0001) and prismatic (101̄0)⟨11̄20⟩ slip systems, which govern the plastic response in this orientation.

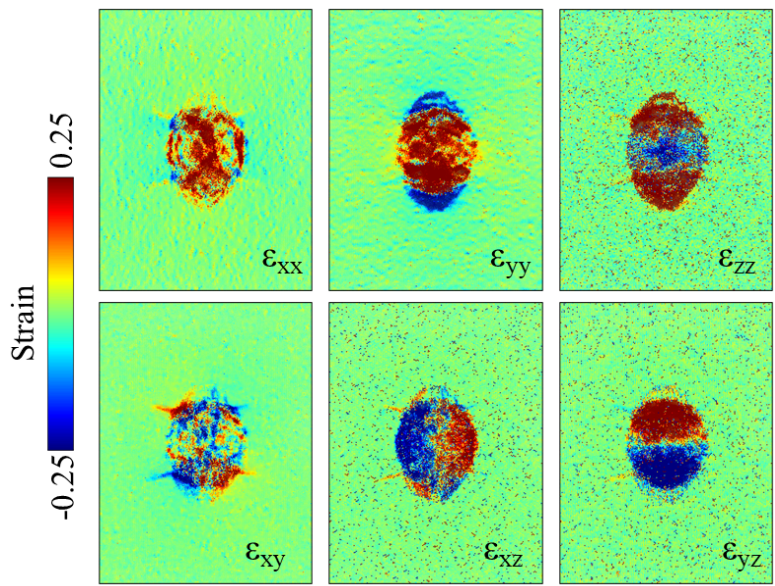

**Fig. 16.** Atomic strain tensor components for [21-10]-oriented Ti after nanoindentation at 10, 300, and 600 K

## 3.3. Dislocation Density Analysis

To quantify the extent of plastic deformation, the dislocation density ρ was computed as a function of indentation depth using the relation:

$$\rho = \frac{\Sigma L_i}{V_D}, \qquad (5)$$

where $L_i$ represents the total length of dislocation segments of type $i$, and $V_D$ is the volume of the plastically deformed region. The latter was approximated by considering a hemispherical plastic zone, with volume:

$$V_D = (2\pi / 3) \times (R_{pl}^3 - h^3), \qquad (6)$$

where $R_{pl}$ denotes the maximum radial extent of dislocation lines from the indentation axis, and $h$ is the indentation depth. This geometrical model accounts for the growth of the plastic zone beneath the indenter and provides a consistent framework for comparing dislocation activity across crystallographic orientations in HCP Ti. The results reported in Table 1 indicate that dislocation density is consistently highest in the [0001] orientation, confirming its greater propensity for dislocation nucleation and accumulation. The prismatic orientations [10$\bar{1}$0] and [$\bar{2}$110] show lower values, in agreement with the observed dual-lobe and asymmetric pile-up morphologies. Increasing temperature generally reduces dislocation density in the prismatic orientations due to enhanced recovery and annihilation, while in the basal orientation [0001], dislocation density increases with temperature, reflecting persistent nucleation activity and retention of defects even at elevated thermal conditions.

**Table 1.** Computed dislocation densities (ρ) presented in $\times 10^{16} m^{-2}$ for [0001], [10$\bar{1}$0], and [$\bar{2}$110] orientations at different temperatures.

| Temperature | [0001] | [10-10] | [2-1-10] |
|---|---|---|---|
| 10K | 0.86 | 0.45 | 0.41 |
| 300K | 1.09 | 0.40 | 0.37 |
| 600K | 1.52 | 0.38 | 0.12 |

These quantitative results are in line with the qualitative observations presented earlier. Orientations that showed broader plastic zones and more diffuse shear strain fields also exhibit lower residual dislocation densities, consistent with enhanced recovery and annihilation. In contrast, the [0001] orientation, where slip activity is more confined and phase transformation contributes to defect stabilization, retains a higher dislocation density across all temperatures. Taken together, the comparison of density values with strain maps and surface morphologies gives a consistent view of how orientation and temperature shape the plastic response of Ti under nanoindentation.

## 4. Conclusion

Our results demonstrate that both temperature and crystallographic orientation decisively shape the nanoscale plastic response of single-crystalline α-titanium under nanoindentation. Load–displacement analysis revealed distinct elastic–plastic transitions, with pop-in events most pronounced for the [0001] orientation and smoother nucleation behavior in the prismatic orientations. Dislocation network analysis confirmed that basal a-type dislocations dominate plasticity, while pyramidal ⟨c+a⟩ slip becomes increasingly active with temperature, enabling strain accommodation along the c-axis. By combining orientation-resolved Schmid factor analysis with shear strain mapping, we established a direct link between stress distribution, slip system activation, and the resulting surface morphologies. Importantly, the [0001] orientation retained the highest dislocation density across all temperatures, reflecting persistent nucleation and limited recovery, whereas prismatic orientations exhibited lower densities due to enhanced annihilation and thermal recovery processes. Taken together, the simulations shed light on how thermally activated mechanisms shape indentation-induced plasticity in titanium and reproduce surface features in close agreement with AFM experiments. These results provide a mechanistic framework for interpreting temperature-sensitive deformation in hcp Ti, offering valuable guidance for the design and optimization of Ti-based materials in structural applications operating across a wide temperature range.

**Acknowledgments**

Research was funded through European Union Horizon 2020 research and innovation program under Grant Agreement No. 857470 and from the European Regional Development Fund under the program of the Foundation for Polish Science International Research Agenda PLUS, Grant No. MAB PLUS/2018/8, and the initiative of the Ministry of Science and Higher Education "Support for the activities of Centers of Excellence established in Poland under the Horizon 2020 program" under Agreement No. MEiN/2023/DIR/3795. The Ministry of Science, Technological Development, and Innovation of the Republic of Serbia, grant No. 451-03-136/2025-03/200023. We gratefully acknowledge Polish high-performance computing infrastructure PLGrid (HPC Center: ACK Cyfronet AGH) for providing computer facilities and support within computational Grant No. PLG/2024/017084. This publication is based upon work from MecaNano COST Action CA21121, for the STMS supported by COST (European Cooperation in Science and Technology).